\documentclass[%
reprint,
 amsmath,amssymb,
 aps,
pra,
notitlepage,
]{revtex4-2}
\usepackage{graphicx}
\usepackage{dcolumn}
\usepackage{bm}
\usepackage{hyperref}
\usepackage{textgreek}
\usepackage{xcolor}
\usepackage[caption=false]{subfig}

\begin{document}
\preprint{APS/123-QED}

\title{Dark solitons in Fabry-P\'erot resonators with Kerr media and normal dispersion}
\author{Graeme N. \surname{Campbell$^{1}$}}
\email{graeme.campbell.2019@uni.strath.ac.uk}
\author{Lewis \surname{Hill$^{1,2}$}}
\author{Pascal \surname{Del'Haye$^{2,3}$}}
\author{Gian-Luca \surname{Oppo$^{1}$}}
\affiliation{$^1$SUPA and Department of Physics, University of Strathclyde, 
Glasgow, G4 0NG, Scotland, UK\\ 
$^2$Max Planck Institute for the Science of Light, 91058 Erlangen, Germany\\ 
$^3$Department of Physics, Friedrich Alexander University Erlangen-Nuremberg, 91058 Erlangen, Germany
}

\begin{abstract}
    Ranges of existence and stability of dark cavity-soliton stationary states in a Fabry-P\'erot resonator with a Kerr nonlinear medium and normal dispersion are determined. The Fabry-P\'erot configuration introduces nonlocal coupling that shifts the cavity detuning by the round trip average power of the intracavity field. When compared with ring resonators described by the Lugiato-Lefever equation, nonlocal coupling leads to strongly detuned dark cavity solitons that exist over a wide range of detunings. This shift is a consequence of the counterpropagation of intracavity fields inherent to Fabry-P\'erot resonators. At difference with ring resonators, the existence and stability of dark soliton solutions are dependent on the size and number of solitons in the cavity. We investigate the effect of nonlocal coupling of Fabry-P\'erot resonators on multiple dark solitons and demonstrate long range interactions and synchronization of temporal oscillations.
\end{abstract}


\maketitle


\section{Introduction}
Temporal cavity solitons (TCS) \cite{coen2016temporal} have attracted significant interest for the generation of broadband optical frequency combs \cite{PasquaziReview18} with many applications in telecommunication \cite{pfeifle2014coherent,pfeifle2015optimally}, spectroscopy \cite{suh2016microresonator,dutt2018chip} and in the fundamental studies of complex dissipative structures. TCS in high Q ring microresonators are now routinely used to produce frequency combs. Light propagation in micro-ring resonators is well described by the longitudinal version of the Lugiato-Lefever equation (LLE) \cite{lugiato1987spatial}. The LLE originally described the transverse, dissipative spatial structures in passive optical systems with diffraction and was later adapted to describe pattern formation along the cavity length \cite{haelterman1992dissipative,lugiato2018lugiato}.

In recent years the generation of bright TCS within Fabry-P\'erot (FP) resonators in the anomalous dispersion regime has been first studied analytically \cite{cole2018theory} and then experimentally demonstrated for continuous wave \cite{wildi2022soliton} and pulsed \cite{obrzud2017temporal} driving. This has resulted in increasing interest in linear resonators for the generation of TCS as the FP geometry can offer additional engineering possibilities when compared to a ring resonator for tailoring dispersion and allowing for greater control over the bandwidth and temporal duration of cavity solitons. Such possibilities include the engineering of the core-cladding index \cite{yang2016broadband}, analogous to the engineering of ring resonator geometry, or the design of the mirror dispersion \cite{wildi2022soliton}.

Here we extend the work of \cite{cole2018theory} to the case of normal dispersion where stable dark cavity-solitons (DCS) are found as opposed to bright ones. We model a FP resonator filled with a Kerr nonlinear medium and investigate the inherent counterpropagation of light under normal dispersion in Section II. We describe first how the FP configuration results in many unstable stationary solutions, which are stable in an equivalent ring resonator in Section III. In spite of this fact, solutions containing moving fronts between bistable states exist and are described in Section IV. It is these moving fronts that can lock with each other and form DCS solutions. DCS stationary solutions are found to be detuning shifted with respect to those in a ring resonator described by the LLE by the average power of the field over a round trip of the cavity. This is a result of an additional nonlocal coupling term exhibited by the FP resulting from counterpropagation of the intracavity fields. To properly elucidate the effects of the shift in detuning, stationary solutions of the FP model and their stability are compared with those of ring resonators with normal dispersion \cite{parra2016dark,parra2016origin}, those of FP with anomalous dispersion \cite{cole2018theory} and those of counterpropagating light in ring-resonators with normal dispersion \cite{campbell2022counterpropagating}. Finally we investigate oscillatory DCS solutions in a FP and discuss the effects of the nonlocal coupling on the oscillating solitons, homogeneous background, and the interaction of two oscillating DCS. In particular, the long range interaction between DCS is capable of synchronizing their oscillations.

\section{The Fabry-P\'erot model}
\begin{figure}
    \centering
    \includegraphics[width = 0.8\linewidth]{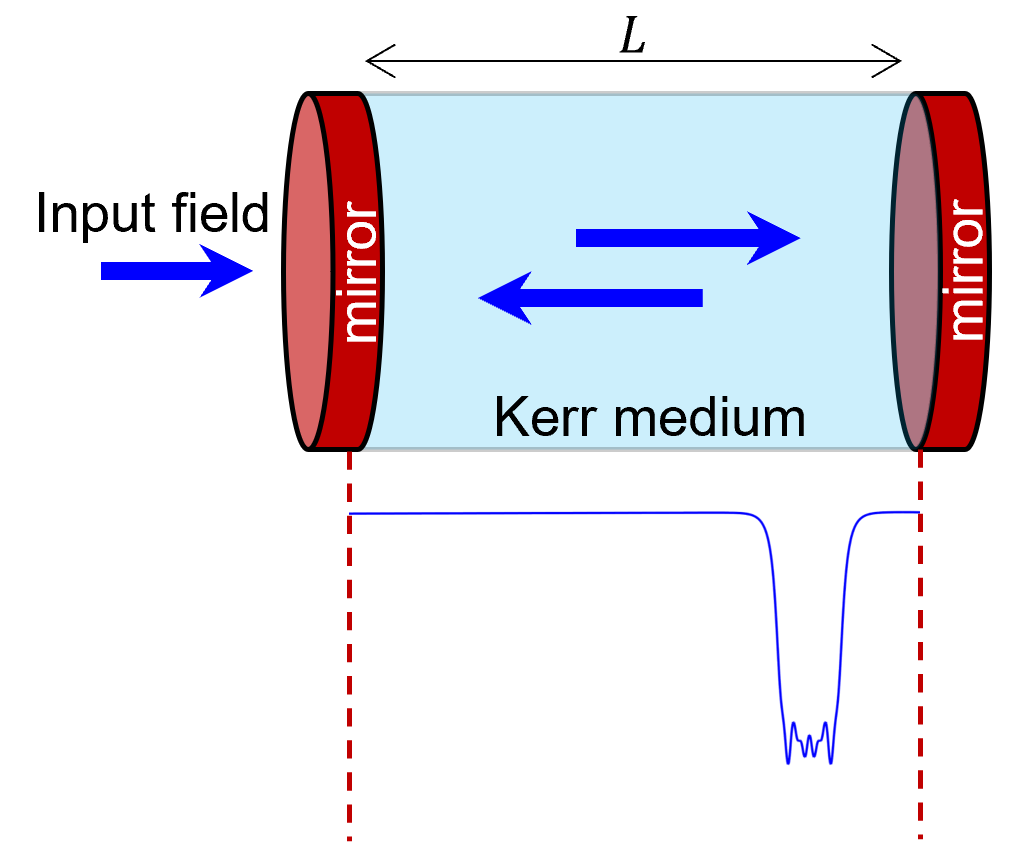}
    \caption{Setup: A Fabry P\'erot resonator filled with a Kerr medium. An input field pump enters the cavity on one side and part of the field is coupled out upon each reflection at the output mirror. An example field power displaying a DCS is shown. }
    \label{fig:Fabry-Perotsetup}
\end{figure}

We consider a high finesse FP resonator composed of highly reflective mirrors and filled with a Kerr medium, see Fig. \ref{fig:Fabry-Perotsetup}. The resonator is driven by linearly polarized light, which is coupled through one of the cavity mirrors into the resonator and the intracavity field is coupled out upon each reflection. This system has been previously studied in the case of anomalous dispersion in \cite{cole2018theory}, where a so called Lugiato-Lefever equation for the Fabry-P\'erot (FP-LLE) was derived. For normal dispersion the dimensionless FP-LLE has the form
\begin{eqnarray}
    \partial_t \psi &=& S - (1 + i\theta)\psi + i( |\psi|^2 + 2\langle |\psi|^2\rangle)\psi - i\partial^2_\zeta\psi\label{eq:FPLLE}
\end{eqnarray}
where $\psi(\zeta,t)$ is the normalized field envelope composed of the combined forward and backward counterpropagating  fields in the FP resonator defined over the extended domain $-L\leq\zeta\leq L$ where $L$ is the dimensionless resonator length, $S$ is the amplitude of the normalized coupled input field, here considered to be real and positive, and $\theta$ is the normalized detuning to the near nearest cavity resonance. Here, $t$ is the `slow time' temporal variable describing the evolution over many round trips of the cavity whilst $\zeta$ is the `fast time' spatial variable describing the evolution over a single round trip of the linear cavity. The term notated with angled brackets represents an integral
\begin{equation}
    \langle |\psi|^2\rangle = \frac{1}{2L}\int_{-L}^L |\psi|^2 d\zeta
\end{equation}
corresponding to the average power of the intracavity field over a round trip of the resonator. This term arises due to the counterpropagation of the fields which is inherent in a FP configuration. Integral terms of this kind are also present in bidirectionally pumped ring resonators where there are instead two coupled equations, one for each counterpropagating field \cite{Kondratiev20,Skryabin:20,campbell2022counterpropagating}, and are the result of rapid phase dynamics of cross-coupling terms due to the fields seeing each other through an average intensity. 

A connection between Eq. (\ref{eq:FPLLE}) and a LLE for a ring resonator system has been previously made in \cite{cole2018theory} and investigated in the context of anomalous dispersion. By writing Eq. (\ref{eq:FPLLE}) as
\begin{eqnarray}
    \partial_t \psi &=& S - (1 + i\theta_\mathrm{eff})\psi + i|\psi|^2\psi - i\partial^2_\zeta\psi \label{eq:LLE}
\end{eqnarray}
we can see that the stationary solutions of Eq. \ref{eq:FPLLE} are also stationary solutions of the LLE with an identical input field $S$ and an effective detuning 
\begin{equation}
    \theta_\mathrm{eff} = \theta - 2\langle|\psi|^2\rangle \label{eq:effectiveLLE}.
\end{equation}
that is shifted by the average power of the intracavity field. The role of this effective detuning in counterpropagation in a ring resonator was was investigated in \cite{campbell2022counterpropagating}. The shift in detuning introduced by the FP configuration results in distinct features of the TCSs as discussed in the following sections.  

\section{Homogeneous states and plateaus}\label{sec:fast time dynamics}

We describe stationary solutions composed of flat increments over the cavity round trip of the FP model in the fast time, the normalized spatial coordinate $\zeta$. We write Eq. (\ref{eq:FPLLE}) for $\partial_t \psi = 0$ as 
\begin{eqnarray}
    \partial_\zeta U &=& \Tilde{V},\quad\partial_\zeta V = \Tilde{U},\nonumber\\
    \partial_\zeta \Tilde{U} &=& - (\theta-\langle U^2+V^2\rangle) U -V +UV^2 + U^3\label{eq:fasttimedynamics}\\
    \partial_\zeta \Tilde{V} &=& - (\theta-\langle U^2+V^2\rangle) V + U +VU^2 + V^3 - S\nonumber
\end{eqnarray}
where $U,V$ are the real and imaginary components of $\psi$. We consider the system of Eq. (\ref{eq:fasttimedynamics}) to evolve in fast time $\zeta$ over a round trip of the cavity and we adopt the terminology of \cite{parra2016dark,doi:10.1137/070698191,haragus2011local,champneys1998homoclinic,colet2014formation}. Under this construction, the fixed points $U_0,V_0$ of Eq. (\ref{eq:fasttimedynamics}) correspond to $\partial_\zeta U = \partial_\zeta V = \partial_\zeta \Tilde{U} = \partial_\zeta \Tilde{V} = 0$. Due to the presence of the nonlocal terms, the fixed points of Eq. (\ref{eq:fasttimedynamics}) depend on the round trip average power of the field and require the full evolution over the round trip of the resonator for their determination. Fixed points corresponding to homogeneous stationary states (HSS) satisfy $\langle|\psi_\text{s}|^2\rangle = |\psi_\text{s}|^2$ and can be found by solving (see \cite{cole2018theory,moroney2022kerr})
\begin{equation}
    H^3 - 2(\theta - 2H)H^2 + \{(\theta - 2H)^2 + 1\}H = S^2\label{eq:HSSFP}
\end{equation}
where $H^2=|\psi_\text{s}|^2=\langle|\psi_\text{s}|^2\rangle$. 

The linear stability analysis of the FP HSSs can be found in \cite{cole2018theory} showing that the middle power HSS in the simultaneous presence of three HSSs is always unstable. The HSSs are fixed points of Eq. (\ref{eq:fasttimedynamics}) when the average power over the round trip is equal to the fixed point power. As such it is not possible for solutions that start and return to a homogeneous state (including front and DCS solutions, see Sections \ref{sec:Switching fronts} and \ref{sec:DCS}) to hang from the HSS as the presence of a nonlocal fast time inhomogeneity would change the average power of the field. Hence the HSS of the FP model are a subset of fixed points which do not support exponentially localized solutions. 

We consider now the fixed points of Eq. (\ref{eq:fasttimedynamics}) corresponding to homogeneous plateaus connected by step functions. These plateau states have uniform power $Y_\pm$ of finite size $1-\Delta$ and $\Delta$, respectively, in the normalized fast time variable $\zeta/(2L)$ and are then different from the HSSs. The round trip average power is given by $\langle |\psi|^2\rangle = \Delta Y_- + (1-\Delta)Y_+$ parameterized by the two bistable plateau powers and the plateau durations with $0<\Delta<1$. The plateau powers are then the solutions of the coupled cubic equations
\begin{eqnarray}
    Y_\pm^3 &-& 2(\theta - 2\Delta Y_- - 2(1-\Delta)Y_+)Y_\pm^2\\ 
    &+& \{(\theta - 2\Delta Y_- - 2(1-\Delta)Y_+)^2 + 1\}Y_\pm = S^2 \, .\nonumber \label{eq:plateauFP}
\end{eqnarray}
For each value of the plateau duration $\Delta$, there are up to nine possible solutions. When restraining to real solutions that satisfy $Y_+ > Y_-$, at most three pairs of $(Y_+,Y_-)$ solutions remain. For each of these pairs, one can evaluate the effective detuning and find that only one pair corresponds to the upper-lower HSS of the corresponding LLE, if these exist. In this way one obtains two plateaus powers for each value of $\Delta$ unless the effective detuning is outside the bistability region of the corresponding LLE. This approximation has been used to great effectiveness to describe and predict stable switching fronts between plateaus in bidirectionally pumped ring resonators \cite{campbell2022counterpropagating}. We will see in Section \ref{sec:Switching fronts} that solutions of plateau connected by switching fronts are dynamical for the FP model while they have large ranges of stability in bidirectional ring resonators. 

The fast time stability of the plateau states can be understood by considering the Jacobian matrix 
\begin{equation}
    J = 
    \begin{pmatrix}
    0 & 0 & 1 & 0\\
    0 & 0 & 0 & 1\\
    V^2 +3U^2-\theta_\mathrm{eff} & -1 + 2UV & 0 & 0\\
    1 + 2UV & U^2 +3V^2-\theta_\mathrm{eff} & 0 & 0
    \end{pmatrix}_{(U_s,V_s)}\label{eq:jac}
\end{equation}
where $U_s$ and $V_s$ are the steady states of Eqs. (\ref{eq:fasttimedynamics}) and $\theta_\mathrm{eff}$ is the effective detuning corresponding to the stationary field. The Jacobian matrix assumes negligible change to the round trip average power and provides eigenvalues of the form
\begin{equation}
    \lambda = \pm\sqrt{(2Y-\theta_\mathrm{eff}) \pm \sqrt{(Y^2-1)} }
    \label{fastevalues}
\end{equation}
where $Y = U_0^2+V_0^2$ is the plateau power of the stationary state $U_s, V_s$ from Eq. (\ref{eq:fasttimedynamics}). The Jacobian (\ref{eq:jac}) has a similar form to the fast time analysis performed for a LLE \cite{parra2016dark}, where $\theta$ replaces the effective detuning and the plateaus correspond to the HSS. The eigenvalues (\ref{fastevalues}) rule the escape from and the approach to plateau states $Y_\pm$ along the stable and unstable manifolds. Of key relevance is the transition that occurs at plateau power $Y = 1$, below which the eigenvalues (\ref{fastevalues}) become complex. The lower power plateau typically exist beneath this threshold, $Y_-<1$, displaying fast time oscillation. This allows for the structurally stable intersection of stable and unstable manifolds of the plateau corresponding to the formation of dark cavity solitons \cite{knobloch2015spatial}. These eigenvalues are essential in the determination of the presence of local oscillations responsible for the existence and stability of dark cavity solitons in Section \ref{sec:DCS}. We note that the higher power plateau always has power $Y_+>1$ and yields four real eigenvalues for the parameters we have considered.

\subsection{Plateau stability in the slow time}

The linear stability of the HSS of the FP has been been previously explored \cite{cole2018theory} but due to the presence of the round trip average term in Eq. (\ref{eq:FPLLE}) the existence and linear stability of a plateau solution depends now on the full field variable along $\zeta$. Here we investigate the linear stability in the slow time of solutions $\psi = \psi_+ + \psi_-$ formed by two plateaus $\psi_\pm$, coexisting on a round trip and joined by two step functions in fast time. We determine the linear stability of these solutions to perturbations of the form 
\begin{equation}
    \psi_\pm = \psi_{\pm,\text{s}} + \left(\epsilon_\pm e^{ik_\pm\zeta+\Omega_\pm t} + c.c. \right) \, .
\end{equation}
By assuming that the wave numbers $k_\pm$ are periodic on their respective plateau the effect of the perturbation on the round trip average becomes $\langle|\psi_\text{s}+\epsilon|^2\rangle = \langle|\psi_\text{s}|^2\rangle$, and we find that the growth rate of the perturbation on the higher power $\Omega_+$ and lower power $\Omega_-$ plateaus are
\begin{eqnarray}
    \Omega_+ &=& -1 \pm \sqrt{4Y_+\theta_\mathrm{eff} -3Y_+^2 \theta_\mathrm{eff}^2 -(4Y_+ - \theta_\mathrm{eff})k_+^2 - k_+^4} \nonumber \\
    \Omega_- &=& -1 \pm \sqrt{4Y_-\theta_\mathrm{eff} -3Y_-^2 \theta_\mathrm{eff}^2 -(4Y_- - \theta_\mathrm{eff})k_-^2 - k_-^4} \nonumber \\
    & & \label{eq:plateaugrowth}
\end{eqnarray}
where $\theta_\mathrm{eff} = \theta - 2\langle|\psi_\text{s}|^2\rangle$ is the effective detuning shifted by the average power of the stationary field. The eigenvalues (\ref{eq:plateaugrowth}) have a similar form to that of the HSS but now with a dependency on the fast time average. The dependence on the round trip average power is explicit for the plateau states due to fast time inhomogeneity over the round trip such that when $Y$ is equal to $\langle|\psi|^2\rangle$ the eigenvalues of Eq. (\ref{eq:plateaugrowth}) reduce to the eigenvalues of the HSS and are comparable to the stability eigenvalues of the LLE \cite{parra2016dark}. The eigenvalues of Eq. (\ref{eq:plateaugrowth}) predict a Turing instability of the lower power plateau solution starting at the threshold $Y=1$ present for $\theta - 2\langle|\psi_\text{s}|^2\rangle\geq 2$. The critical wave number associated with the maximum growth is given by $k_Y^2 = 2(\theta - 2\langle|\psi|^2\rangle - 2Y_-)$. The HSS of the FP also displays a Turing instability starting at the threshold $H=1$ when $\theta - 2H \geq 2$ \cite{cole2018theory}, and exhibits the critical wave number $k_H^2 = 2(\theta - 4H)$. Generally, the average intensity of the dark plateau solutions is much larger than the lower power HSS, and as such the critical wave number of the plateau is much smaller than that associated to the HSS, i.e. $k_Y<k_H$.

If we now consider a time dependent homogeneous perturbation to the plateaus of the form
\begin{equation}
    \psi_\pm = \psi_{\pm,\text{s}} + \epsilon_\pm'(t)
\end{equation}
we can investigate the linear stability of the plateau solutions to perturbations which change the average power of the field in the slow time $\partial^2_\zeta\psi = 0$. We derive a Jacobian matrix of the linearized plateaus system as can be seen in Appendix \ref{app:Jacplat} and evaluate the eigenspectrum of this Jacobian matrix numerically.
\begin{figure}[!ht]
    \centering
    \includegraphics[width = 1\linewidth]{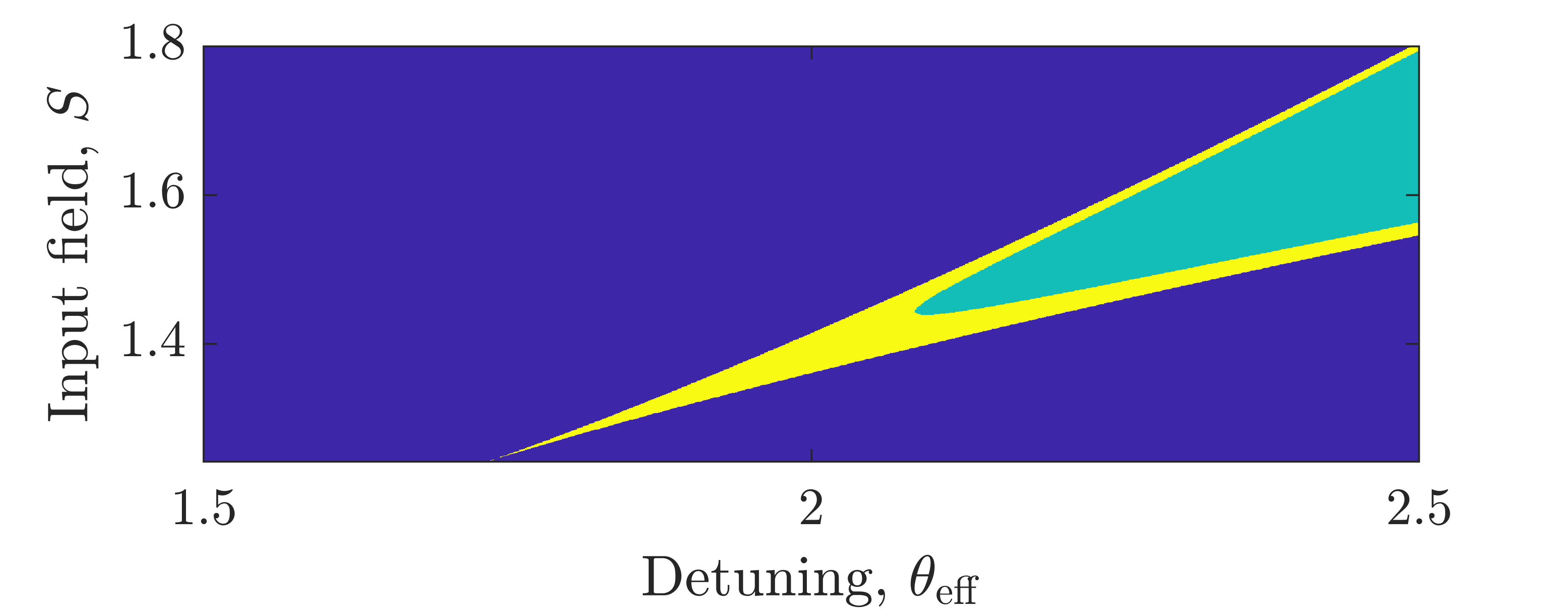}
    \caption{Stability of coexisting plateaus in the parameter space of shifted detuning, $\theta_\mathrm{eff}$, and input field, $S$, for a fixed length of the lower power plateaus $\Delta = 0.2$. The blue regions correspond to single stable HSS of the FP model, the green regions to two stable plateaus coexisting on the round trip of the resonator, and the yellow region to unstable plateau solutions.}
    \label{fig:eigenvaluesPlateau}
\end{figure}

In Fig. \ref{fig:eigenvaluesPlateau} the stability of plateau states to perturbations of the average power is reported in the parameter space $(\theta_\mathrm{eff},S)$ when the lower power plateau has size $\Delta = 0.2$. This allows for a direct comparison with the solutions of bistable HSS as seen for a ring resonator \cite{parra2016dark}. We find that near the onset of optical bistability, plateau solutions in the FP configuration are unstable, as depicted by the yellow zone of Fig. \ref{fig:eigenvaluesPlateau}. The green zone shows stable plateaus, and the blue indicates solutions of a single stable HSS. We find parameter regimes in which there is a bistability of the HSS of the FP model but there are no stable solutions of coexisting plateaus. In regions where plateau solutions are unstable, the system cannot support exponentially localized structures which approach them such as switching fronts (see next section) and indeed we do not observe the formation of stable dark cavity-solitons in this regime.

\section{Switching fronts and their dynamics}\label{sec:Switching fronts}

\begin{figure}[]
    \centering
    \includegraphics[width = 0.95\linewidth]{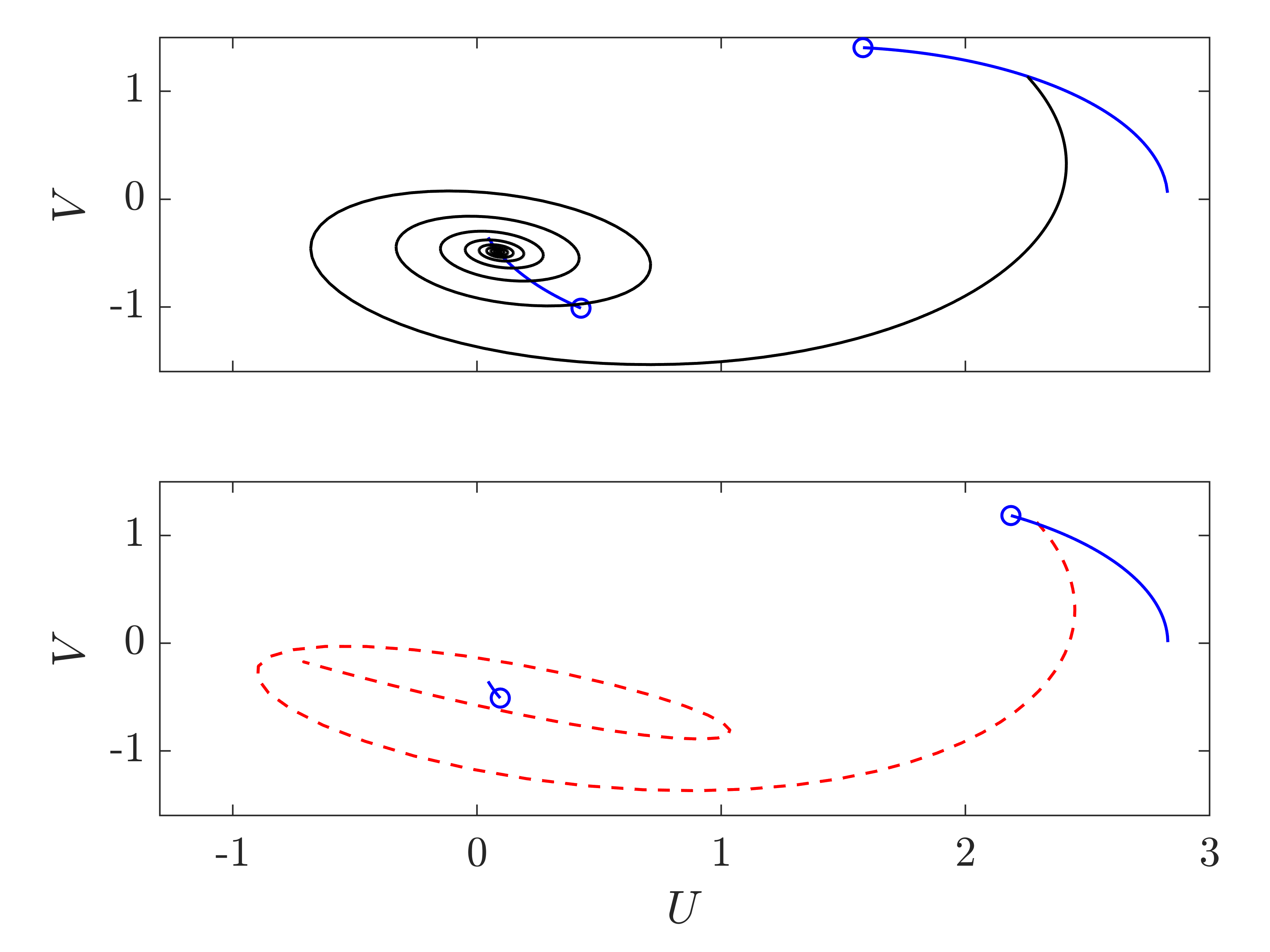}
    \caption{Examples of SF heteroclinic (black line top panel) and DCS homoclinc (red dashed line bottom panel) trajectories in the $V$ vs $U$ plane of Eqs. (\ref{eq:fasttimedynamics}) where $U,V$ are the real and imaginary parts of the intracavity field respectively. Solid blue lines correspond to upper and lower plateaus of different size $\Delta$. Circles correspond to HSS.}
    \label{fig:heteromo}
\end{figure}

In the previous section we have studied the existence and stability of plateau solutions connected with a step function. These are approximations to the solutions of Eq. (\ref{eq:fasttimedynamics}) that approach flat solutions along the fast time variable $\zeta$. There are two kinds of such solutions: heteroclinic orbits, i.e. trajectories that connect two separate fixed points of Eq. (\ref{eq:fasttimedynamics}), and homoclinic orbits corresponding to a trajectory that leaves and returns to the same fixed point of Eq. (\ref{eq:fasttimedynamics}). The former corresponds to switching fronts (SF) solutions while the latter are here associated to cavity solitons: bright (dark) cavity solitons in the case the fixed point corresponds to the high (low) power plateau. As the boundary conditions of a FP resonator are periodic, SF solutions exist as pairs with opposite orientation in the cavity. Taken together, they form a heteroclinic cycle. DCS solutions described later in this paper are themselves composed of oppositely oriented SFs which interact and lock with each other through local oscillations close to the lower power fixed point. An example of SF (heteroclinic) and DCS (homoclinic) in the $(U,V)$ plane are presented in Fig. \ref{fig:heteromo}. These trajectories are anchored to the plateau solutions discussed in Section \ref{sec:fast time dynamics} for a given value of the distance $\Delta$. The blue solid lines in Fig. \ref{fig:heteromo} correspond to the family of plateau solutions when changing $\Delta$ while the circles mark the positions of the HSS.

\begin{figure*}
    \centering
    \includegraphics[width = .5\linewidth]{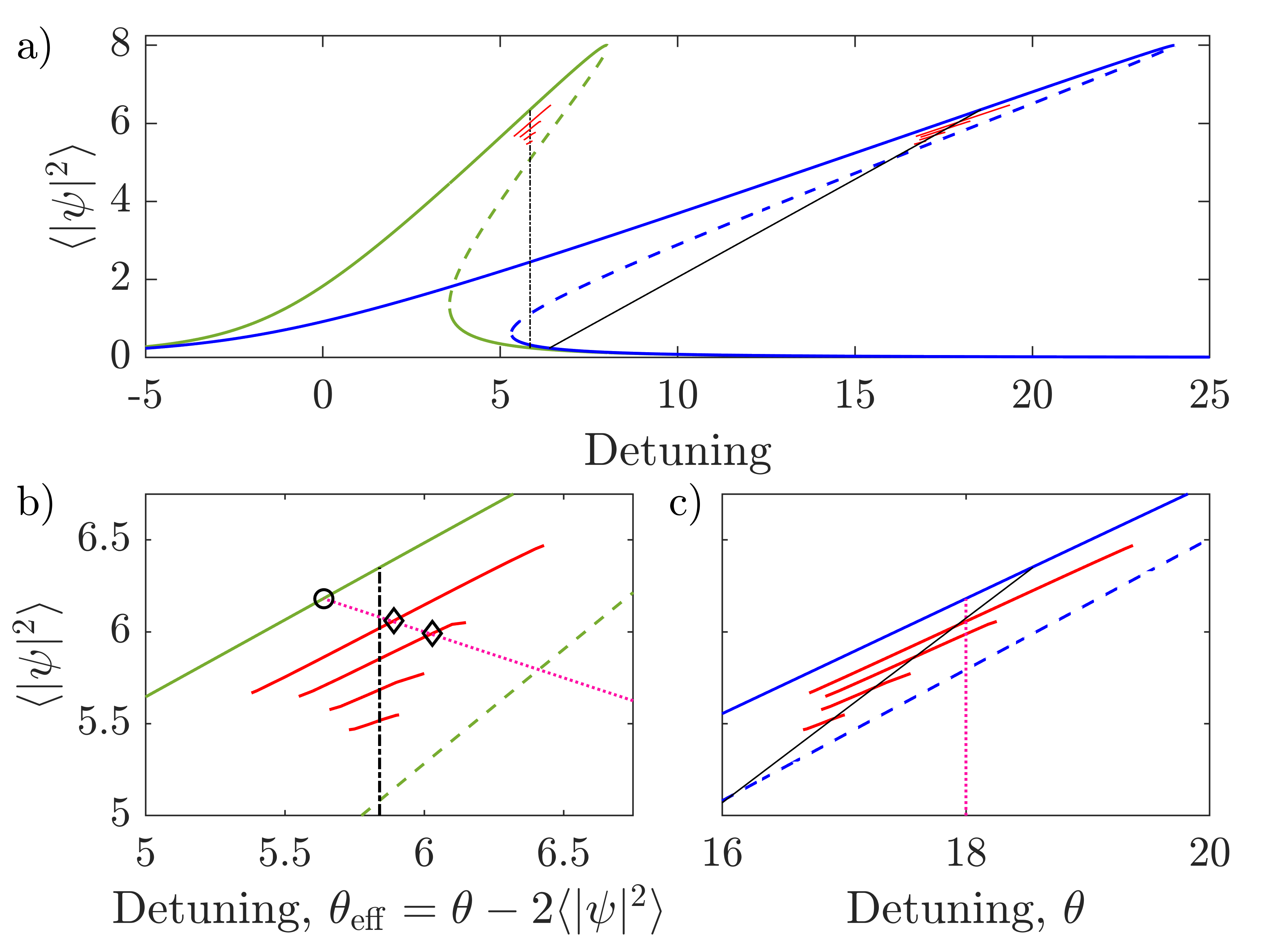}\includegraphics[width = .5\linewidth]{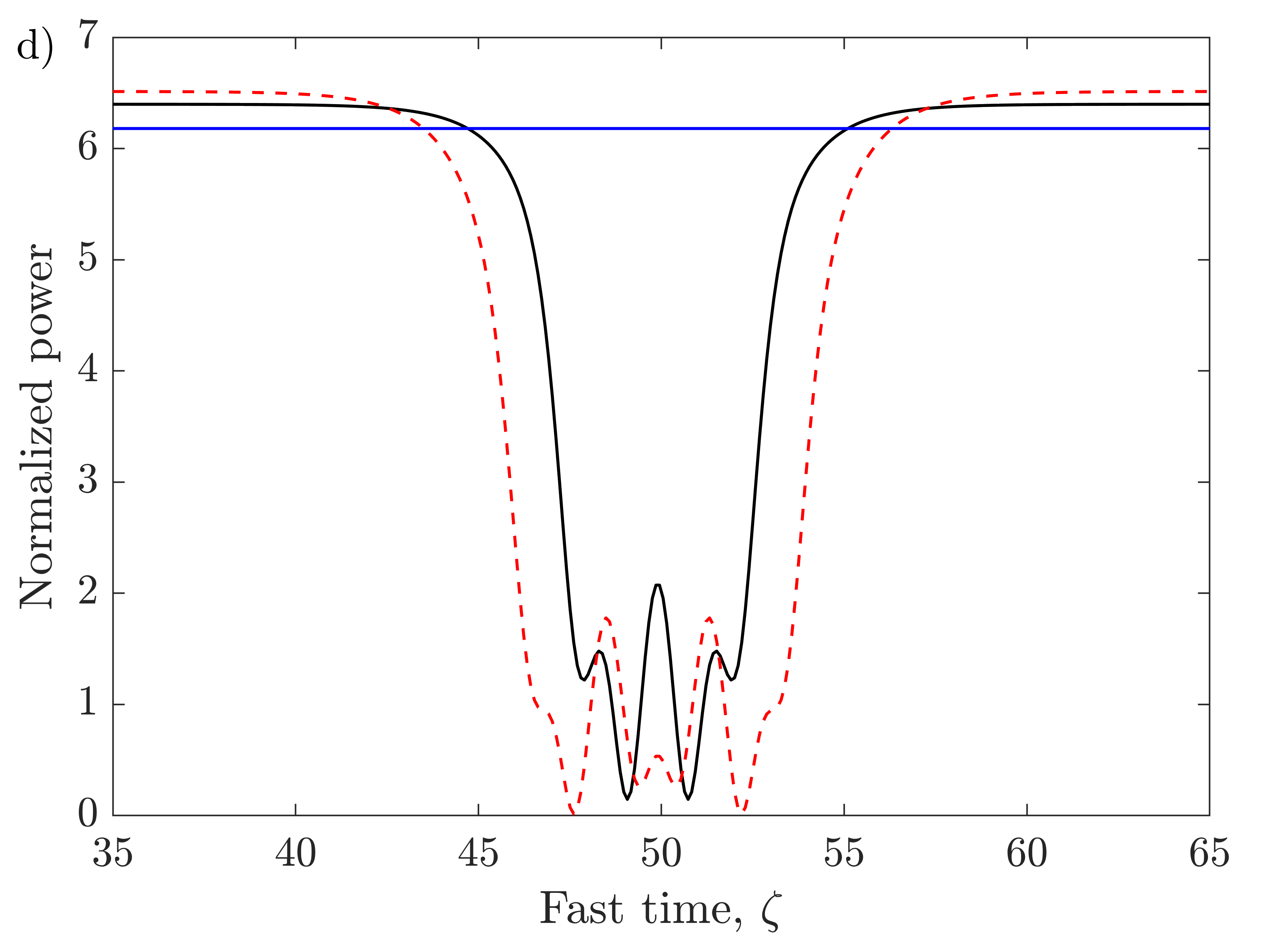}\caption{(a) Solutions of the Fabry-P\'erot model and equivalent ring-resonator model for input field $S = 2\sqrt{2}$ with FP length $2L=100$, and ring circumference $L=100$. HSS of the FP (ring-resonator) model correspond to blue (green) lines. Dashed curves correspond to unstable HSS. Stationary SF solutions are plotted by using their round trip average power for the FP model (black line) and for the ring-resonator model (black dot-dashed line). Stable dark solitons solutions of different size correspond to the red lines and form branches of distinct width. (b) Solutions of the ring-resonator model plotted with respect to the ring-resonator detuning $\theta_\mathrm{eff}$ which are related to the FP solutions through the effective detuning, Eq. (\ref{eq:effectiveLLE}), and corresponding to a zoom of window (a). (c) Solutions of the FP model plotted with respect to the detuning and corresponding to a zoom of window (a). (d) Power profile of bistable stationary dark solitons for parameters $S=2\sqrt{2}$, $\theta = 18, \, 2L=100$ and corresponding to the two diamonds in (b). The solid blue line in (d) corresponds to the highest power HSS marked with a circle in (b). 
    }
    \label{fig:HSSandDarksolitons}
\end{figure*}
\begin{figure}
    \centering
    \includegraphics[width = 1\linewidth]{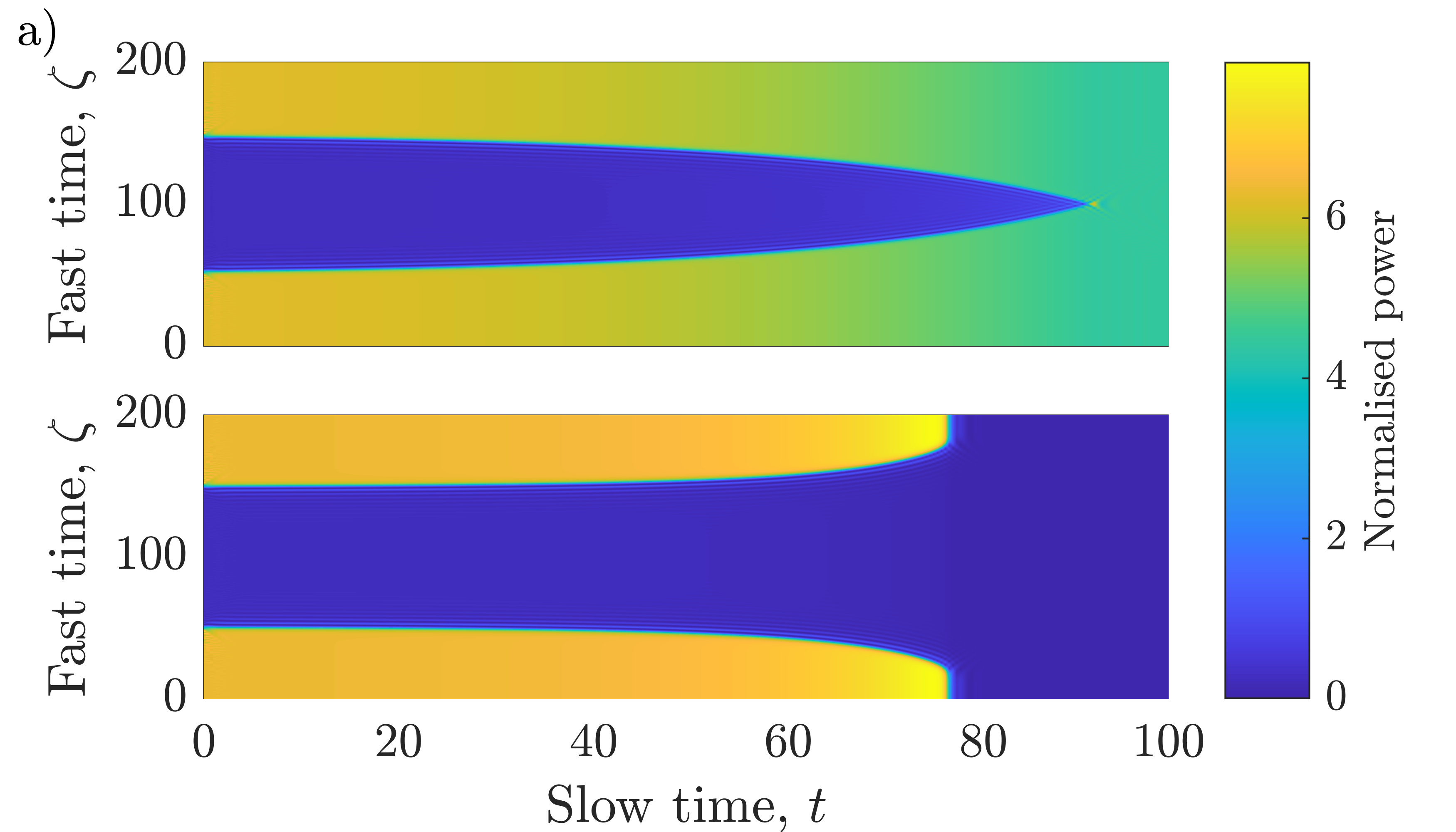}
    \includegraphics[width = 1\linewidth]{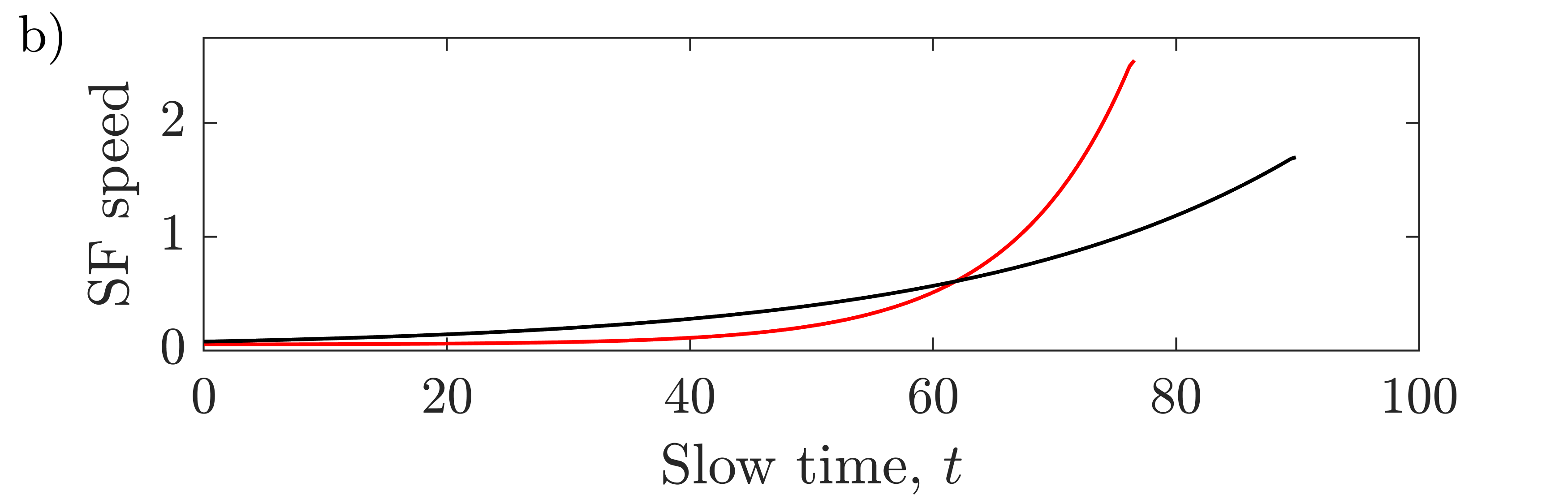}
    \caption{a) SFs moving away from the spatial point where their velocity changes direction at separation of $\Delta\approx 0.5$ for parameters $S=2\sqrt{2}$, $\theta = 12.5$. The upper panel (lower panel) shows the evolution of two SFs from a square wave initial condition with slightly smaller (larger) SF distance than the turning point. In both cases the SFs move away from the starting location, annihilate each other and collapse to one of the bistable HSSs. b) SF velocities corresponding to the upper panel (black line) and lower panel (red) of a).}
    \label{fig:unstableSFs}
\end{figure}

We now discuss the stationary states and dynamics of SF (heteroclinic) solutions found for the FP model and make comparison with similar solutions found in ring resonators. The HSSs of the FP model are plotted in Fig. \ref{fig:HSSandDarksolitons}a as blue curves. The HSSs of an equivalent ring-resonator model are plotted as green curves. When compared to the case of a single field in a ring-resonator, it becomes clear that the effect of the shift in the detuning is most prominent for high power solutions, resulting in a large shift of the peak resonance. SF solutions of Eq. (\ref{eq:FPLLE}) present as oppositely oriented pairs and are generally dynamical, moving with identical speed in opposite directions. The velocity of SF solutions depend on of the average power of the fields and can display a turning point where the velocity of both SFs change sign. An example evolution is shown in Fig. \ref{fig:unstableSFs} for fixed parameters from an initial condition near a velocity turning point at $\Delta\approx 0.5$. In the top (bottom) panel of Fig. \ref{fig:unstableSFs} the SF initial condition has separation slightly narrower (wider) than at the velocity turning point. We can see that the SF solutions either move towards each other until they annihilate (top panel), or move away from each other until the effective detuning is shifted beyond the bistability region of the equivalent LLE (bottom panel). 

The velocity turning points of the SF of the FP resonator are plotted in Fig. \ref{fig:HSSandDarksolitons}a as solid black lines, and correspond to a specific value of $\Delta$ where the effective detuning correspond to the Maxwell point of the LLE (see dash-dot black line in Fig. \ref{fig:HSSandDarksolitons}a). The Maxwell point of the LLE corresponds to the unique value of detuning $\Theta_\mathrm{MP}$ for the chosen parameters where the velocity of a non interacting SF is zero. This then constitutes a stationary state of the FP model with a value of $\Delta$ providing an effective detuning that is equal to $\Theta_\mathrm{MP}$. Due to the dependence of the effective detuning on the average power of the field there exist a single separation of SF corresponding to a single turning point for the FP. The turning point can be then located semi-analytically given that we know $\Theta_\mathrm{MP}$ as shown for example for bidirectionally pumped ring resonator in \cite{campbell2022counterpropagating}. The direction of the SF motion is then determined by the effective detuning being greater than $\Theta_\mathrm{MP}$ (Fig. \ref{fig:unstableSFs}a upper panel) or smaller than $\Theta_\mathrm{MP}$ (Fig. \ref{fig:unstableSFs}a lower panel).

For counterpropagating light in ring resonators, there exists an abundance of stable and robust light plateau stationary states composed of non-interacting SFs that form in one field, while the counterpropaging field is flat in profile \cite{campbell2022counterpropagating}. The ring resonator system is described by two equations with similar form to Eq. (\ref{eq:FPLLE}), one each for the forward and backward counterpropagating fields. It is interesting to note that for counterpropagation in a ring-resonator the SF with a given initial separation move towards (instead of away from) the velocity turning point \cite{campbell2022counterpropagating}. The difference in direction of SF motion when compared with the FP model is due to distinct backward and forward field profiles present for counterpropagation in ring-resonators, such that the effective detuning of each field depends on the power of the counterpropagating field and not explicitly on its own average power. Unlike counterpropagating fields in ring-resonators, the stability of SF solutions in a FP configuration depends critically on its own average power resulting in SFs solutions that always move away from the turning point (see Fig. \ref{fig:unstableSFs}). This explains why we do not observe the rich phenomenology of stable SFs solutions of the counterpropagation in ring-resonators \cite{campbell2022counterpropagating} and why only DCS are achieved through the locking mechanism originating from the interaction of SF through their oscillating tails as described in the next section.

Finally, we note that although SF heteroclinic solutions are not stationary in an FP configuration, the calculations of the plateau solutions in Section \ref{sec:fast time dynamics} are still worth mentioning. During the motion displayed in Fig. \ref{fig:unstableSFs}, the solution progresses through the plateau solutions of different size $\Delta$ instant by instant until the two SF interact with each other. During the SF motion the cusp point of the heteroclinic trajectory of the top panel of Fig. \ref{fig:heteromo} moves along the blue curves that are the plateau solutions of Eq. (\ref{eq:plateauFP}). If the initial $\Delta$ is larger than that corresponding to the velocity turning point, the cusp of the heteroclinic trajectory on the right hand side of the the top panel of Fig. \ref{fig:heteromo} moves leftward along the blue solid line until it reaches the HSS marked by the circle. If the initial $\Delta$ is smaller than the velocity turning point, the cusp of the heteroclinic trajectory on the right hand side of the the top panel of Fig. \ref{fig:heteromo} moves rightward along the blue solid line until the SF annihilate and the system collapses to the lower HSS (circle closer to the origin of the axes).

\section{Dark cavity solitons}\label{sec:DCS}
For a FP resonator with normal dispersion, we observe the formation of DCS steady states. Such states are composed of two SFs that lock with each other because of the interaction of spatial oscillations present close to the lower plateau, as shown for single field ring-resonators in \cite{parra2016dark} and in optical parametric oscillators \cite{oppo1999domain,oppo2001characterization}. Bistable DCS stationary solutions of Eq. (\ref{eq:FPLLE}) are shown in Fig. \ref{fig:HSSandDarksolitons}d for parameter values $S=2\sqrt{2},\, \theta=18, \, 2L=100$. The two dark soliton solutions have different widths corresponding to distinct cycles of the locked oscillatory tails. Such solitons can be obtained by a perturbation of the HSS of suitable width. 

It can be seen that the plateau power is different from that of the HSS (blue line in Fig. \ref{fig:HSSandDarksolitons}d) and also for each DCS of different widths. As was discussed in section \ref{sec:fast time dynamics}, the plateau power of an exponentially localized state depends on the average power of the field over a round trip such that solitons of different width display different plateau power due to their different average power. The DCS of Fig. \ref{fig:HSSandDarksolitons}d correspond to the diamonds in \ref{fig:HSSandDarksolitons}b. The plateau solutions from which a DCS hangs from correspond to the points on the solid blue curve of the lower panel of Fig. \ref{fig:heteromo} with the value of the shifted detuning, $\theta_\mathrm{eff}$, of the corresponding distance $\Delta$. Note that the bistable DCS and the HSS of the FP model are distributed along the line $\langle|\psi|^2\rangle = -\theta_\mathrm{eff}/2 + \theta/2$ as shown in Figs. \ref{fig:HSSandDarksolitons}b and \ref{fig:HSSandDarksolitons}c as a pink dotted line, respectively.

The effect of the shift in detuning discussed throughout this paper can be understood by making a comparison between the DCS of normal dispersion described here and the FP bright solitons in the anomalous dispersion regime \cite{cole2018theory}. Due to the large average power of DCS solutions, there is a much larger shift in detuning when compared with the bright solitons in the anomalous dispersion case. DCS solutions are also much further detuned than cavity solitons in an equivalent ring resonator.  
\begin{figure}
    \centering
    \includegraphics[width = 1\linewidth]{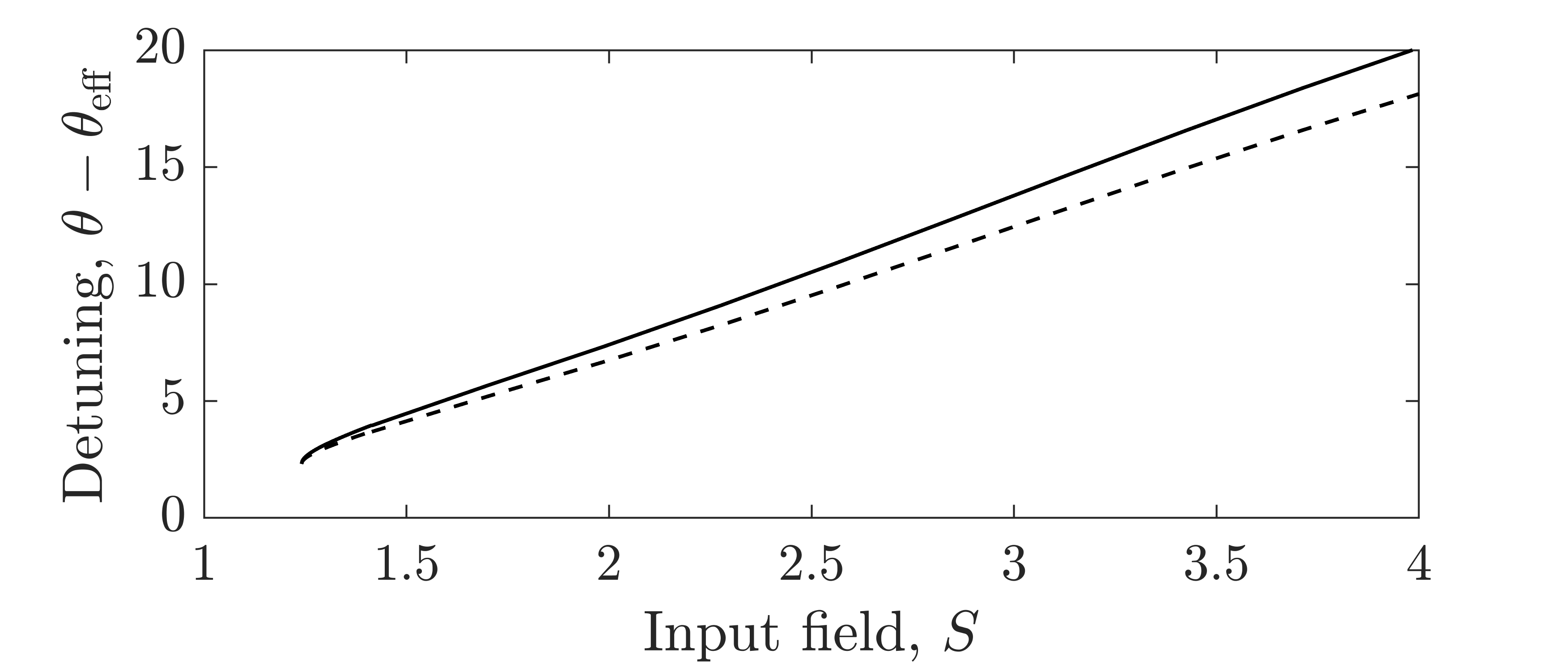}
    \caption{Approximate difference in detuning between dark soliton solutions of the FP and equivalent ring-resonator models with respect to the common input field. The black line corresponds to the SF turning point when $\Delta = 0$ for the FP model. The dashed line corresponds to the SF turning point when $\Delta = 0.1$.}
    \label{fig:Solitonlocation}
\end{figure}
To see this, we plot in Fig. \ref{fig:Solitonlocation} the difference in detuning values between the FP resonator and an equivalent ring-resonator model $\theta_\mathrm{eff}$, and corresponding to the location of DCS for given input powers $S$. The location of a DCS is approximated by using the SF turning point line (an example of which is shown in Fig. \ref{fig:HSSandDarksolitons}a) intersecting either the high power HSS (solid black line in Fig. \ref{fig:Solitonlocation}) or by selecting a point along this line corresponding to a specific DCS size $\Delta$. The dashed black line in Fig. \ref{fig:Solitonlocation} shows solutions in which SFs are stationary with separation $\Delta = 0.1$, the approximate size of the DCS of Fig. \ref{fig:HSSandDarksolitons}d). These lines were calculated using the numerical fit of the Maxwell point of the ring resonator model from \cite{campbell2022counterpropagating}. We see that as the input power is increased the shift becomes larger. Furthermore, the range of detunings where DCS exist is much larger for the FP model when compared to an equivalent ring-resonator system. For example, the longest DCS solution branch for an FP resonator spans $16.75<\theta<19.37$, see Fig. \ref{fig:HSSandDarksolitons}c, whilst DCS in the equivalent ring-resonator system are present only in the range $5.38<\theta_\mathrm{eff}<6.43$, approximately $2.5$ times smaller.  

\begin{figure}
    \centering
    \includegraphics[width = 1\linewidth]{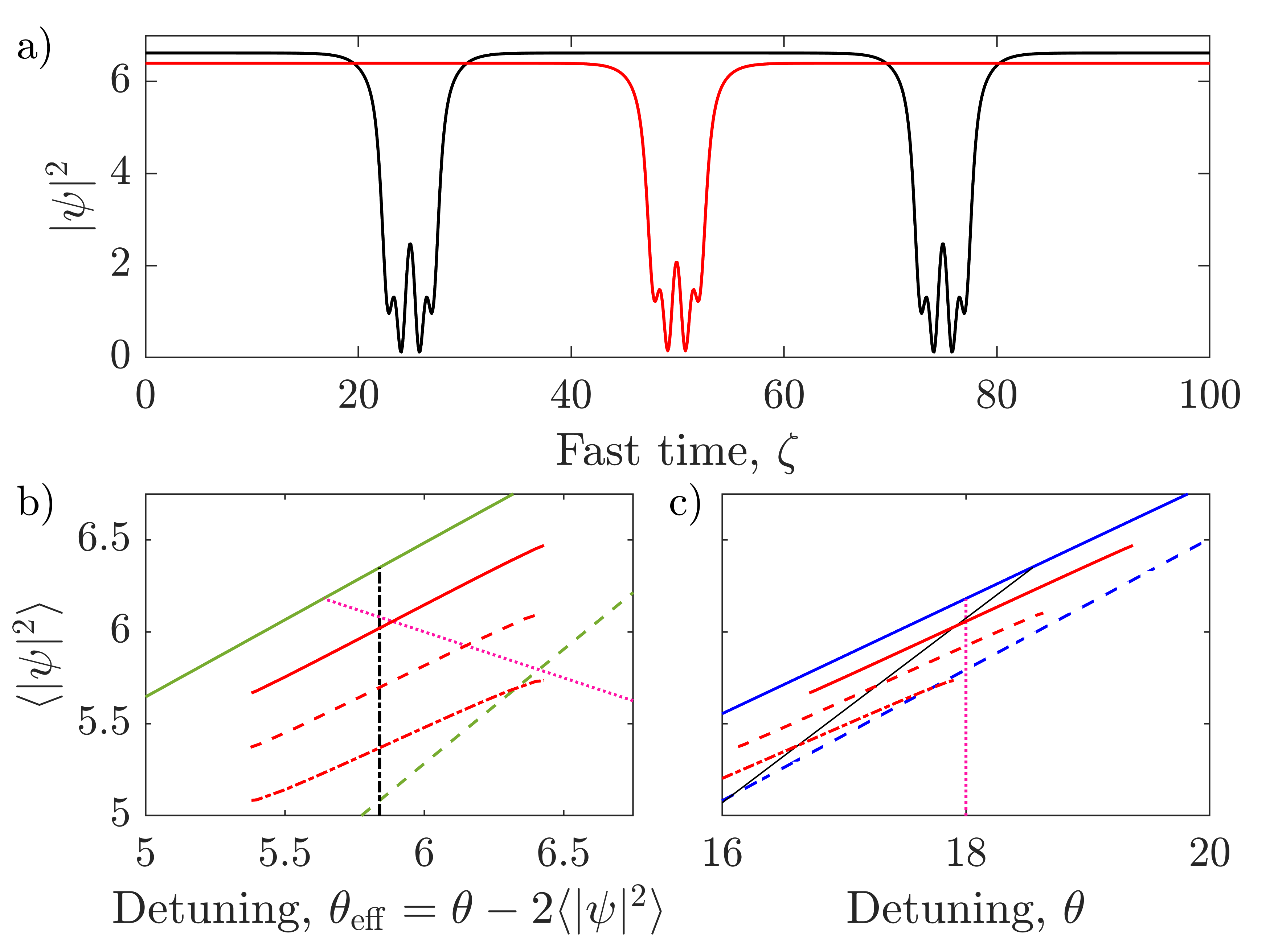}
    \caption{a) Soliton states containing a single soliton, red line, and two solitons, black line for $S=2\sqrt{2}$,$\theta = 18$. b)-c) FP solutions plotted over detuning (or shifted detuning) with respect to their average power. HSS are plotted as blue, b), green c), the black curves corresponds to the SF turning point, and dark soliton solutions are plotted as red curves of one, solid line, two, dashed line, and three, dot-dashed line solitons present in the cavity. The pink dotted line correspond to the solutions with $\theta = 18$.}
    \label{fig:multisoliton}
\end{figure}
The average power of the field is also affected by the number of solitons present in the cavity such that the existence and stability of the solitons depend also on the number of solitons in the cavity. Fig. \ref{fig:multisoliton}a shows two bistable DCS solutions for $S = 2\sqrt{2}$, $\theta = 18$ corresponding to one or two solitons in the cavity. We see that the presence of an additional soliton modifies the low power oscillations and the plateau power. In Figs. \ref{fig:multisoliton}b and \ref{fig:multisoliton}c we plot the DCS solutions as red curves with one soliton, solid, two solitons, dashed, and three solitons, dot-dashed, present in the cavity with an identical number of fast-time oscillations. With each additional soliton in the cavity, the average power of the field decreases. As such, solution branches containing a large number of solitons appear at lower values of detuning. For detuning $\theta = 18$, input field $S = 2\sqrt{2}$ and $2L = 100$, we find that states of one and two solitons are possible, but that further perturbations of the system will not allow the formation of additional solitons, leading instead to the destruction of pre-existing solitons.

\section{Oscillatory dynamics}

\begin{figure}
    \centering
    \includegraphics[width = 1\linewidth]{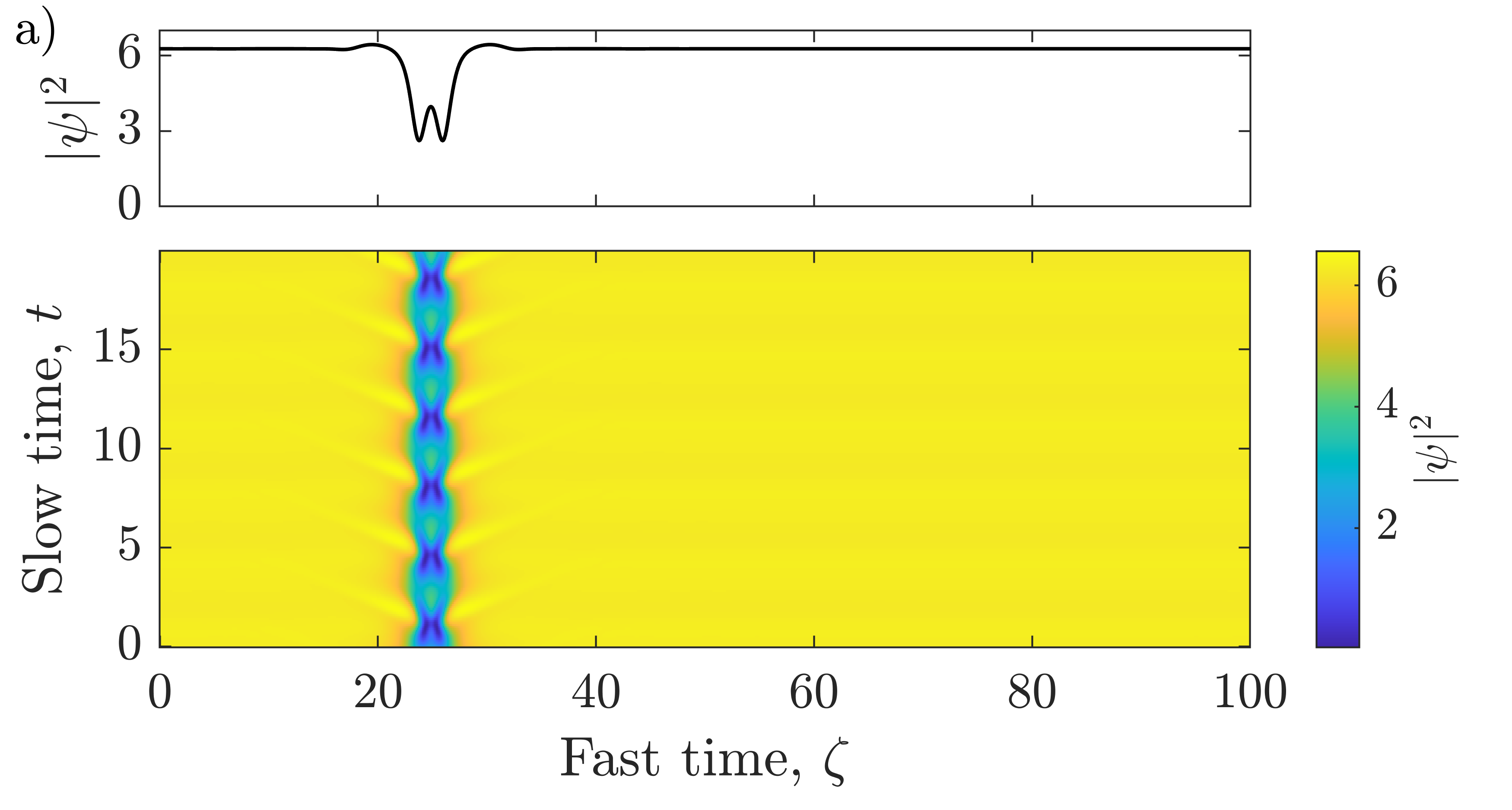}
    \includegraphics[width = 1\linewidth]{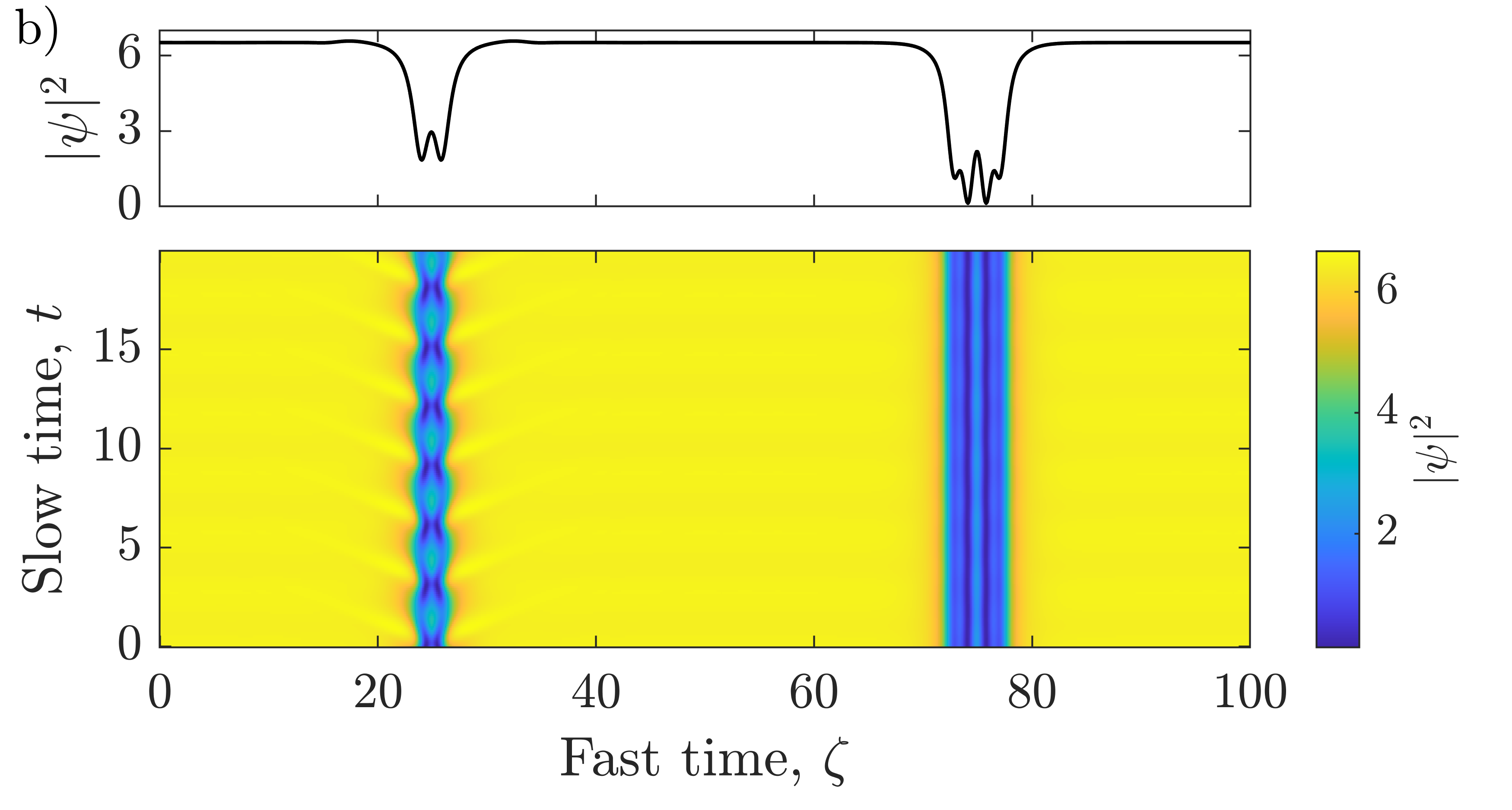}
    \includegraphics[width = 1\linewidth]{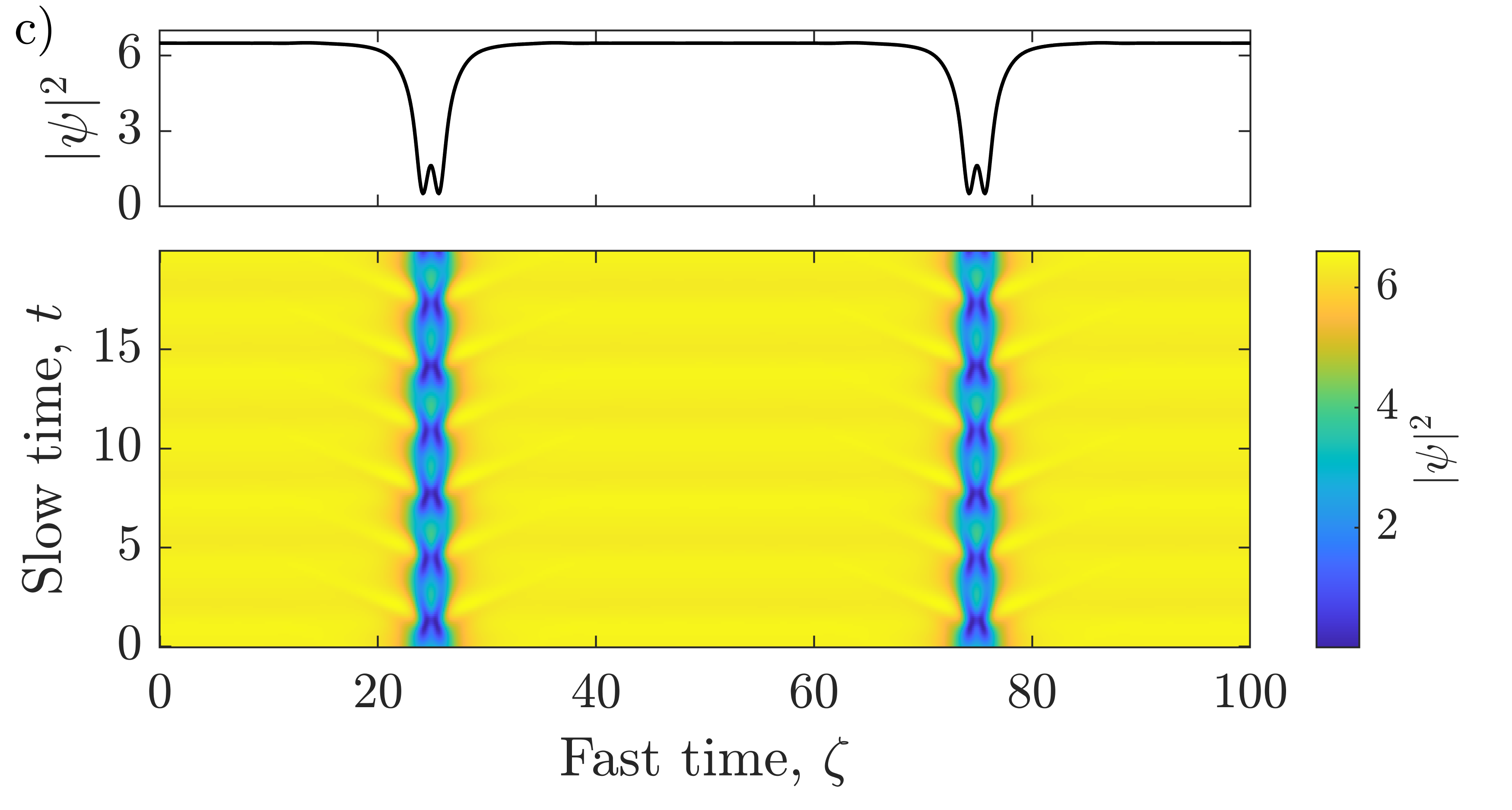}
    \caption{Oscillating solution a) one oscillating soliton, b) one oscillating soliton and one `stationary' soliton, c) two synchronized oscillating solitons (displayed in the upper panels at $t=20$), all for identical parameters $\theta = 18$, $S = 2\sqrt{2}$. The lower panels display the slow time evolution.}
    \label{fig:breathing solitons}
\end{figure}

\begin{figure}
    \centering
    \includegraphics[width = 1\linewidth]{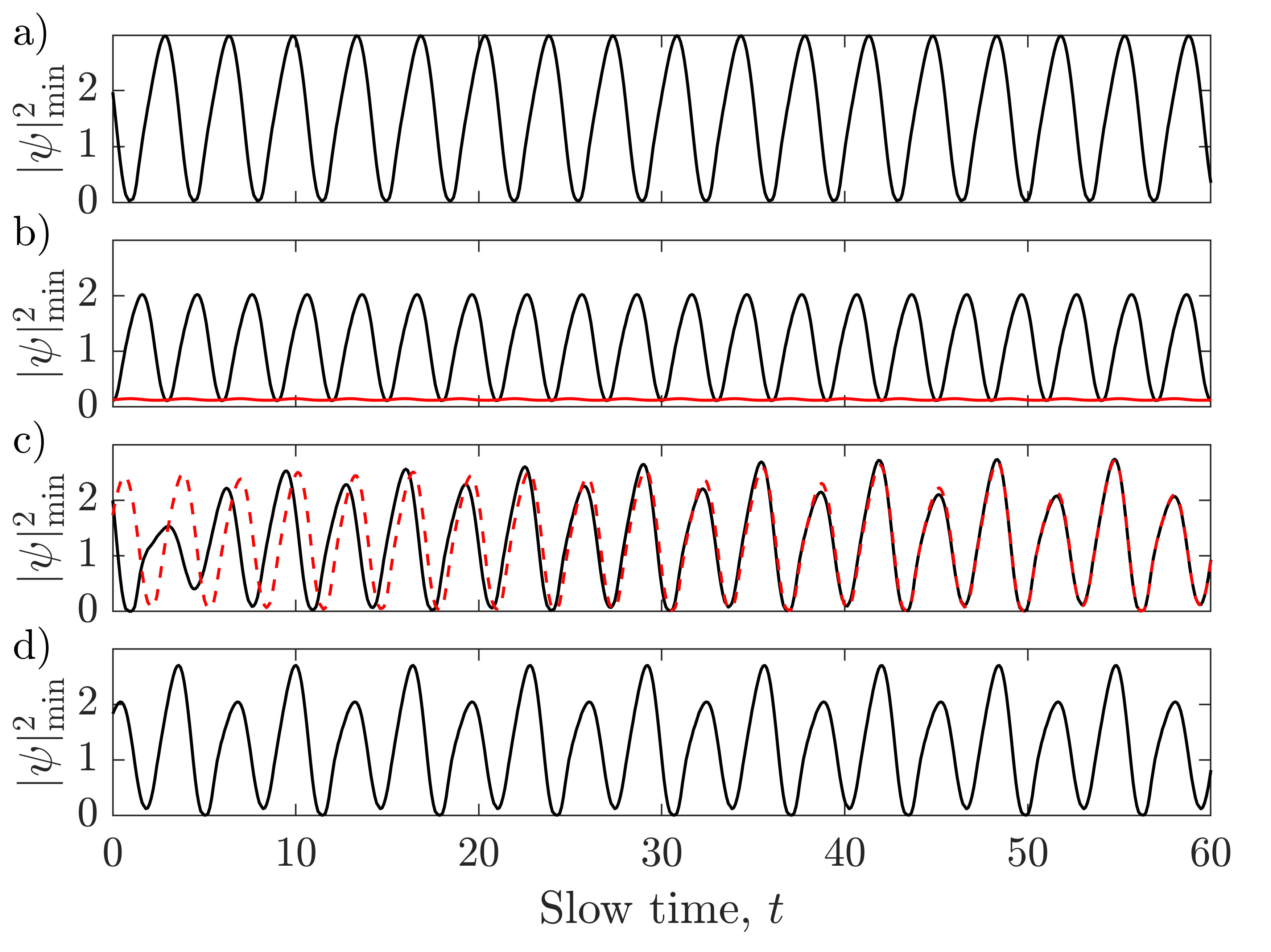}
    \caption{Oscillating solution with a) one oscillating soliton in a FP cavity of length $2L=100$, b) one oscillating soliton and one locked soliton in a FP cavity of length $2L=100$, c) two oscillating solitons in a FP cavity of length $2L=100$, d) one oscillating soliton in a FP cavity of length $2L=50$, all for identical parameters $\theta = 18$, $S = 2\sqrt{2}$.}
    \label{fig:solitonstracePeriod}
\end{figure}
We now investigate the dynamics of DCS states above a temporal instability of Eq. (\ref{eq:FPLLE}). For the case of a ring-resonator, dynamical instabilities of DCS solutions due to Hopf bifurcations result in local oscillations of the soliton \cite{parra2016dark}. In what follows we demonstrate the effects of the nonlocal coupling of Eq. (\ref{eq:FPLLE}) on the dynamics of oscillatory soliton solutions. 

In Fig. \ref{fig:breathing solitons}a we show an example of an oscillatory solution of the FP model with input field $S=2\sqrt{2}$ and detuning $\theta=18$. We can see that the temporal dynamics of the system is not confined to the soliton but it extends to an oscillation of the background plateau too. As the soliton moves through its limit cycle, there is a change in the average power of the field. Hence, due to the nonlocal self interaction term, regions of the cavity far from the soliton exhibit oscillatory dynamics with identical period to that of the soliton. These oscillations are small due to the small change in average power originating from the oscillation of the soliton. A trace of the minimum power of the oscillating soliton is plotted in Fig. \ref{fig:solitonstracePeriod}a, showing a single period of oscillation. In Fig. \ref{fig:breathing solitons}b we introduce an oscillating dark soliton to an initial stationary solution containing a stable stationary dark soliton shown in Fig. \ref{fig:HSSandDarksolitons}d. The presence of the oscillating soliton induces small temporal oscillations in the plateau power and tiny oscillations in the peaks of the pre-existent DCS. The trace of the minimum power is plotted in \ref{fig:solitonstracePeriod}b in black for the oscillating soliton, and in red for the stable soliton. We note that the oscillation period has decreased form Fig \ref{fig:breathing solitons}a. In Fig. \ref{fig:breathing solitons}c, we show the evolution of two synchronized breathing solitons. Here we see a larger oscillation amplitude of the plateau power than in the single soliton case, due to the larger change in the average power of the field resulting from the second oscillating soliton. In Fig. \ref{fig:solitonstracePeriod}c we plot the trace of the minimum power of the two oscillating solitons, seen in \ref{fig:breathing solitons}c, starting from an unsynchronized initial condition. As the system evolves over slow time the soliton phases begin to overlap, resulting in full synchronization. The resulting dynamics has now experienced a period doubling with respect to the single oscillating soliton. In Fig. \ref{fig:solitonstracePeriod}d, we plot the evolution of a single soliton with half the cavity length, $2L=50$, of the previous examples. We see that the dynamics of this single soliton is identical to the synchronized dynamics of the two soliton examples with $2L=100$ of Fig. \ref{fig:breathing solitons}c and \ref{fig:solitonstracePeriod}c. In general we find that the dynamics of $N$ well separated solitons in a cavity of length $L$ synchronizes towards the dynamics of a single soliton in a cavity of length $L/N$.

We find that well separated solitons, located such that they do not interact through the local dynamics at the tails, experience phase dependent interaction through the nonlocal coupling. We note that oscillating solitons of the normally dispersive LLE interact locally through their tails and as such do not exhibit synchronization when well separated in a long cavity \cite{parra2016dark}. In the FP model, the change in average power of the field during an oscillation of a single oscillating soliton is small, and as such the change in the power of the background plateau is also small. By increasing the length of the FP cavity, we can reduce the effects of soliton oscillation on the average power such that we can approach the LLE dynamical and independent solutions. The dynamics of the FP model is most distinct from the ring-resonator LLE when the cavity length is small or the number of oscillating solitons is large displaying long range interactions leading to synchronization.

\section{Conclusions}

\begin{figure}
    \centering
    \includegraphics[width = 1\linewidth]{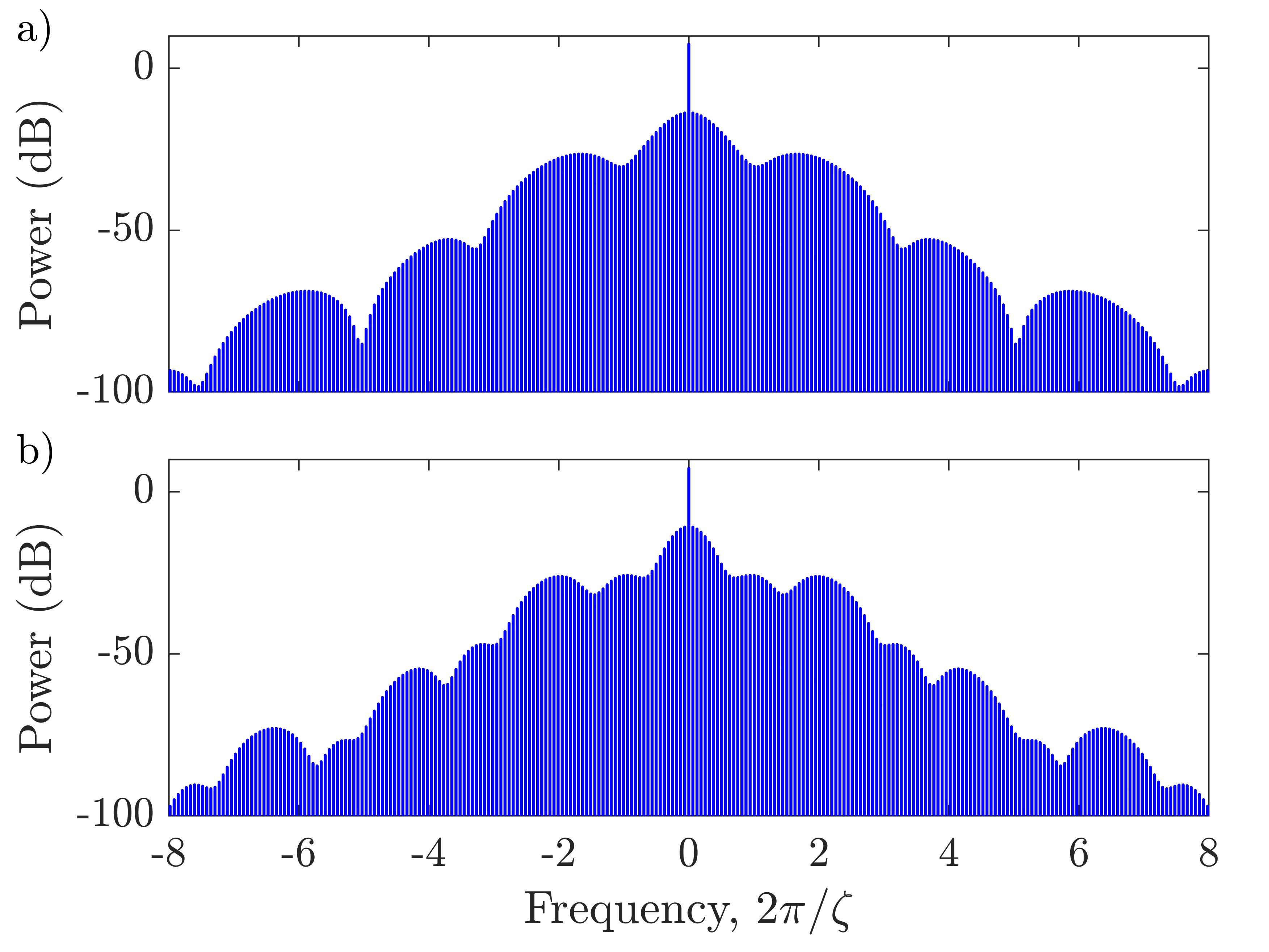}
    \caption{Bistable frequency combs for parameter $S=2\sqrt{2}$, $\theta = 18, 2L=100$. These spectra correspond to the dark solitons shown in Fig. \ref{fig:HSSandDarksolitons}d with lesser a) and greater b) width.}
    \label{fig:darksolitonspectra}
\end{figure}

We have presented bistability, plateaus, switching fronts and dark soliton states in an Fabry-P\'erot model with normal dispersion. Through the use of the effective detuning we demonstrated analogies and differences of these solutions with the stationary states of the ring-resonator case, i.e. the LLE. The homogeneous stationary states of the Fabry-P\'erot model have a one to one correspondence to the stationary states of a ring resonator described by the LLE, with identical input field and a detuning that is shifted by the average power of the intracavity field over a single round trip. We investigated the linear stability of solutions formed by two plateaus that are different from the homogeneous stationary states and connected by step functions. We identified a region of instability near the onset of optical bistability of the homogeneous stationary states. When compared with counterpropagating light in a ring resonator \cite{campbell2022counterpropagating}, it turns out that plateau solutions in a Fabry-P\'erot cavity with normal dispersion are generally unstable. Their stability analysis however is still useful since we can find Turing instabilities of the low power plateaus and determine their critical wave number. Such wave number is much smaller than those corresponding to Turing instabilities of the homogeneous stationary state. It is this wave number that rules the local oscillations close to the lower plateau when two switching fronts lock to form a stable dark soliton. Due to the shift in detuning, a peculiar feature of dark soliton solutions in the Fabry-P\'erot model is that they do not connect to the homogeneous stationary statuses (as they do in the ring-resonator model), but instead to plateau solutions, which are dependent on the size (average power) of the soliton. 

For normal dispersion, the average power of the field, and hence the shift in detuning, is comparatively large for dark solitons when compared with the bright solitons of the anomalous dispersion case. The large shift results in the dark soliton solution branches to be elongated, occupying a larger domain of detuning values than an equivalent ring-resonator configuration, and found in strongly detuned regimes.

Finally, we investigated the effects of nonlocal coupling on dynamics of oscillatory solitons. We find that the nonlocal coupling induces temporal oscillations to the homogeneous background power. In the presence of two oscillating solitons, we observe phase dependent interaction, resulting in synchronization and modifications of their oscillation periods. The resulting synchronized dynamics approach that of a single soliton with half the cavity length. It was found that the dynamics of the system approached that of an equivalent ring-resonator in the limits of a long cavity and small soliton number (small change in average power).

The predictions presented in this paper were obtained for parameter values that are realistically feasible in existing experimental realizations \cite{wildi2022soliton}. When these systems operate at normal dispersion we expect to see the formation of dark solitons steady states in experimental verifications. The Fabry-P\'erot system allows for additional design consideration of the cavity properties allowing for control over frequency comb generation. Example frequency combs of the Fabry-P\'erot resonator are shown in Fig. \ref{fig:darksolitonspectra} which correspond to the dark solitons of Fig. \ref{fig:HSSandDarksolitons}d and show distinct modulations on the combs, corresponding to the number of low intensity spatial oscillations of the dark soliton \cite{nazemosadat2021switching}. The engineering of frequency combs can be useful in applications in precision spectroscopy, LIDAR, and channel generation for telecom systems.



\vfill

\section*{Acknowledgements}
This research was supported by funding from the EPSRC DTA Grant No. EP/T517938/1. P.D. acknowledges support by the H2020 European Research Council (ERC) (756966,Counterlight), the Marie Sklodowska-Curie Innovative Training Network (MSCA) (812818, Microcombs), and the Max Planck Society. L.H. acknowledges funding provided by the CNQO group within the Department of Physics at the University of Strathclyde, and the ``Saltire Emerging Researcher" scheme through the Scottish University's Physics Alliance (SUPA) and provided by the Scottish Government and Scottish Funding Council.

\appendix
\begin{widetext}

\section{Appendix}\label{app:Jacplat}
Here we determine the linear stability of plateau solutions in slow time with $\partial^2_\zeta \psi = 0$. We do so by approximating the intracavity field as a step function $\psi = \psi_+ + \psi_-$ composed of higher power $\psi_+$ and lower power $\psi_-$ plateau solutions such that the average power of the field may be written as $\langle |\psi|^2\rangle = \Delta|\psi_-|^2+(1-\Delta)|\psi_+|^2$ where $\Delta$ is the size of the lower power plateau normalized to the round trip. Hence the higher and lower power plateaus evolve according to the two coupled equations
\begin{eqnarray}
    \partial_t \psi_\pm &=& S - (1 + i\theta)\psi_\pm + i( |\psi_\pm|^2 + 2(\Delta|\psi_-|^2 + (1-\Delta)|\psi_+|^2))\psi_\pm
\end{eqnarray}
The linear stability of the plateau solutions can be understood by finding the eigenspectrum of the Jacobian matrix
\begin{equation}
    J = 
    \begin{pmatrix}
    -1 - A_+ - 4(1-\Delta)U_+V_+ & B_+ - 4(1-\Delta)V_+^2 & -4\Delta U_-V_+ & -4\Delta V_+V_-\\
    -C_+ + 4(1-\Delta)U_+^2 & -1 + A_+ + 4(1-\Delta)U_+V_+ & 4\Delta U_+U_- & 4\Delta U_+V_-\\
     -4(1-\Delta) U_+V_- & -4(1-\Delta) V_+V_- & -1 - A_- - 4\Delta U_-V_- & B_- - 4\Delta V_-^2 \\
     4(1-\Delta) U_+U_- & 4(1-\Delta) U_-V_+ & -C_- + 4\Delta U_-^2 & -1 + A_- + 4\Delta U_-V_- 
    \end{pmatrix}\label{eq:Jacplat}
\end{equation}
where $U_\pm,V_\pm$ are the real and imaginary components of the plateaus and $A_\pm = 2U_\pm V_\pm$, $B_\pm = \theta_\mathrm{eff} - U_\pm^2 - 3V_\pm^2$, $C_\pm = \theta_\mathrm{eff} - 3U_\pm^2 - V_\pm^2$, where $\theta_\mathrm{eff} = \theta - 2\Delta(U_-^2 + V_-^2) - 2(1-\Delta)(U_+^2 + V_+^2)$ is the effective detuning defined in Eq. (\ref{eq:effectiveLLE}). This formulation allows us to investigate the stability of LLE stationary states for the equivalent FP resonator. Numerical evaluation of the eigenspectrum of Eq. (\ref{eq:Jacplat}) yields the results shown in Fig. \ref{fig:eigenvaluesPlateau} and are discussed in Section \ref{sec:fast time dynamics}. 

In the limit $\Delta \rightarrow 0$ we can evaluate the eigenvalues of Eq. (\ref{eq:Jacplat}) as 
\begin{eqnarray}
    \lambda &=& -1\pm\sqrt{A_-^2 - B_-C_-}\label{eq:eigenplateauHOMO1}\\
    \lambda &=& -1\pm\sqrt{(A_++4U_+V_+)^2 - (B_+-4V_+^2)(C_+-4U_+^2)}\label{eq:eigenplateauHOMO2}
\end{eqnarray}
Eigenvalues (\ref{eq:eigenplateauHOMO1}) have an identical form to the eigenvalues of the homogeneous stationary states and hence they have a negative real part for the low power plateaus. Instability of the plateau solutions is therefore due to Eq. (\ref{eq:eigenplateauHOMO2}).
\end{widetext}

\end{document}